\documentclass[%
 reprint,
superscriptaddress,
a4paper,
longbibliography,
amsmath,amssymb,
aps,prl,
numbers,
]{revtex4-1}

\usepackage[pdftex]{hyperref,color,graphicx}
\usepackage{amsfonts,amssymb,amsmath,mathrsfs}
\usepackage[separate-uncertainty=true]{siunitx}
\usepackage[english]{babel}
\usepackage[utf8]{inputenc}
\usepackage[T1]{fontenc}
\usepackage{xspace}
\usepackage[normalem]{ulem}

\newcommand{\figref}[2]{\hyperref[#1]{\getrefnumber{#1}(#2)}}

\begin{document}
\title{Transverse Anderson localization of surface plasmon polaritons}

\author{Z. Cherpakova}
\affiliation{Physikalisches Institut, Rheinische Friedrich-Wilhelms-Universit\"at Bonn, Nu\ss{}allee 12, 53115 Bonn, Germany.}
\author{Felix Bleckmann}
\affiliation{Physikalisches Institut, Rheinische Friedrich-Wilhelms-Universit\"at Bonn, Nu\ss{}allee 12, 53115 Bonn, Germany.}
\author{T. Vogler}
\affiliation{Physikalisches Institut, Rheinische Friedrich-Wilhelms-Universit\"at Bonn, Nu\ss{}allee 12, 53115 Bonn, Germany.}
\author{S. Linden}
\affiliation{Physikalisches Institut, Rheinische Friedrich-Wilhelms-Universit\"at Bonn, Nu\ss{}allee 12, 53115 Bonn, Germany.}



\begin{abstract}
We investigate the effect of disorder on the propagation of surface plasmon polaritons in arrays of evanescently coupled dielectric loaded surface plasmon polariton waveguides. Diagonal disorder is implemented by randomly varying the heights of the waveguides. Real-space as well as Fourier-space images of the surface plasmon polariton intensity distribution in the waveguide arrays are recorded by leakage radiation microscopy. With these techniques we experimentally demonstrate the transverse localization of surface plasmon polaritons with increasing disorder.
\end{abstract}

\maketitle

The propagation of waves in disordered media plays an important role in different branches of physics, e.g., solid state physics, acoustics, and optics. While in the multiple scattering regime, wave transport is in general of diffusive character, in static three dimensional (3D) systems with strong disorder, there can be a phase transition and wave transport comes to a complete halt \cite{anderson1958absence}. This phenomenon is known as Anderson localization and is induced by the enhanced back-scattering interference effect~\cite{lee1985disordered}. 
Signatures of Anderson localization of light in 3D have been experimentally observed in strongly scattering semiconductor powders~\cite{wiersma1997localization}. In particular, the measured transmission coefficient decreased exponentially with the sample thickness. However, it was later pointed out that this observation could be also explained by classical diffusion combined with reasonable amount of absorption~\cite{scheffold1999localization}.

In recent years, photonic lattices and coupled waveguide arrays have become important model systems to study the evolution of waves in complex structures~\cite{christodoulides2003discretizing}. In particular, they allow to simulate the characteristics of quantum systems that are governed by a tight-binding Hamiltonian. By carefully engineering the propagation constant of the individual waveguides as well as their mutual couplings, one can explore a rich set of quantum–optical analogies~\cite{longhi2009quantum}. 
Transverse Anderson localization of light was realized experimentally in 2D disordered photonic lattices \cite{schwartz2007transport}.
Furthermore, the effect of different types of disorder in 1D was studied using evanescently coupled waveguide arrays \cite{martin2011anderson,lahini2008anderson,modugno2009exponential,lahini2009observation,stutzer2013superballistic}. So far, the large majority these experiments have been restricted to dielectric waveguide structures. 

\begin{figure}[tt]
\centering\fbox{\includegraphics[width=\linewidth]{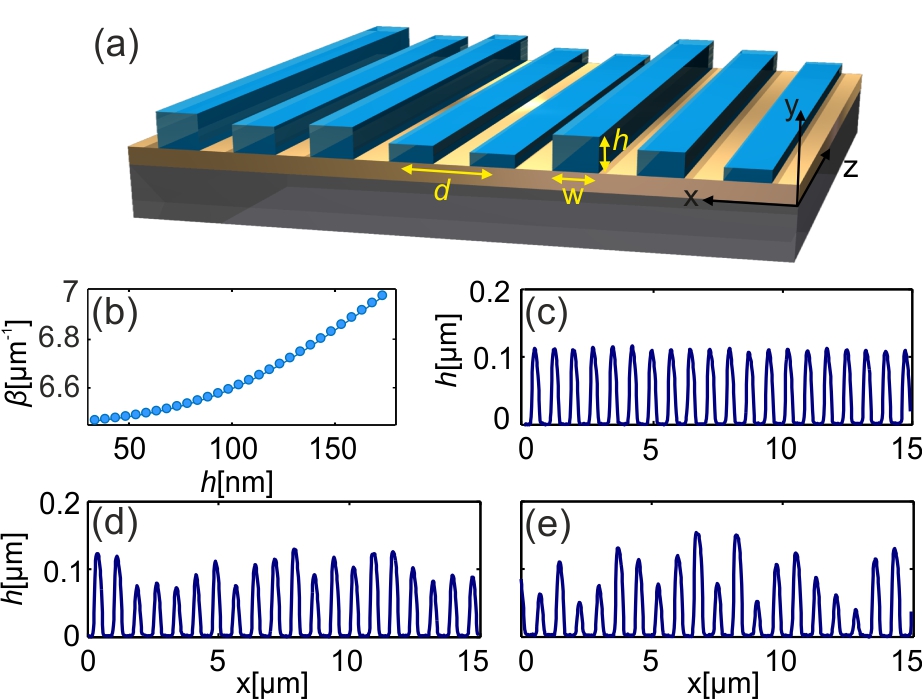}}
\caption{(a) Sketch of the DLSPPW array. (b) Real part of the propagation constant $\beta$  as a function of DLSPPW height~h. (c)~AFM scan of a waveguide array with equal heights ($\Delta\beta/C=0.16$). (d) Typical disordered waveguide array with $\Delta\beta/C=1.2$ (e), and waveguide array with $\Delta\beta/C=1.8$.}
\label{fig:array}
\end{figure}

In this letter we report on the propagation of surface plasmon polaritions (SPPs) in disordered plasmonic waveguide arrays. Diagonal disorder is introduced during fabrication in a controlled way by randomly varying the heights of the waveguides within predefined limits while keeping their separation constant (see scheme in Fig.~\ref{fig:array}(a)).
Using both real- and momentum-space leakage radiation microscopy~\cite{drezet2008leakage} allows us to track the transition from ballistic transport to transverse Anderson localization of SPPs with increasing disorder.

Arrays of evanescently coupled dielectric loaded surface plasmon polariton waveguides (DLSPPWs)~\cite{Holmgaard:2007,Grandidier:2008} are fabricated on top of a cover slip coated with $5\,\rm nm$ chromium and $60\,\rm nm$ gold using negative-tone gray-scale electron beam lithography~\cite{bleckmann2013manipulation,block2014bloch}. The DLSPPWs consist of poly(methyl methacrylate) (PMMA) ridges with a width $w=255\,\rm nm$. 
The height $h$ of each DLSPPW is controlled by the applied electron dose.
This allows us to adjust individually the corresponding real part of propagation constant $\beta$, which is a monotonous function of the DLSPPW height~$h$ (see Fig.~\ref{fig:array}(b)). 
In doing so, the suitable range of heights is constrained by the requirement that all DLSPPWs must support one guided SPP mode.
For all arrays, the center to center distance between neighboring DLSPPWs is chosen to be $d=750\,\rm nm$.
Thus to a first approximation, the coupling constant $C$ between neighboring DLSPPWs remains constant and the ratio $\Delta\beta/C$ can be used as a measure for the degree of disorder~\cite{lahini2008anderson}.

Fig.~\ref{fig:array}(c) presents an atomic force microscopy (AFM) scan of an array without intentional disorder. The DLSPPWs have a mean height $h_{0}=105\, \rm nm$ and a fabrication related height variation of $\delta h=\pm 5\,\rm nm$. 
Finite element calculations show that a single DLSPPW with these geometrical parameters and a free-space wavelength of $\lambda=980\,\rm nm$ supports only one guided SPP mode with a real part of propagation constant $\beta\pm\frac{1}{2}\Delta\beta=6.65\pm\SI{0.02}{\micro\metre}^{-1}$.
With $C=\SI{0.25}{\micro\metre}^{-1}$ (see below), the corresponding disorder parameter is given by $\Delta\beta/C=0.16$.
AFM scans of arrays with two different degrees of diagonal disorder are shown in Fig.~\ref{fig:array}(d) and (e).  
The chosen height variations give rise to the disorder parameters
$\Delta\beta/C=1.2$ and $\Delta\beta/C=1.8$, respectively.
For both degrees of disorder, 15 different realizations are fabricated.

SPPs are excited by focusing a TM-polarized laser beam with $\lambda=980\,\rm nm$ on a grating coupler, deposited on top of the central waveguide. The height of this waveguide is the same in every array to ensure identical excitation conditions. As the SPPs propagate along the waveguides, they coherently couple to radiating modes in the glass substrate. The leakage radiation \cite{drezet2008leakage} as well as the transmitted laser beam are collected with a high numerical aperture oil immersion objective (NA=1.49). Then the transmitted laser beam is removed by Fourier filtering. Real space leakage radiation micrographs with the intensity distribution $I(x,z)$ are recorded by imaging the sample plane onto a sCMOS camera (Andor Zyla).
The corresponding momentum-space intensity distribution $I(k_x,k_z)$ is obtained by imaging the back-focal plane of the oil immersion objective.

\begin{figure}[ttt]
\centering
\fbox{\includegraphics[width=\linewidth]{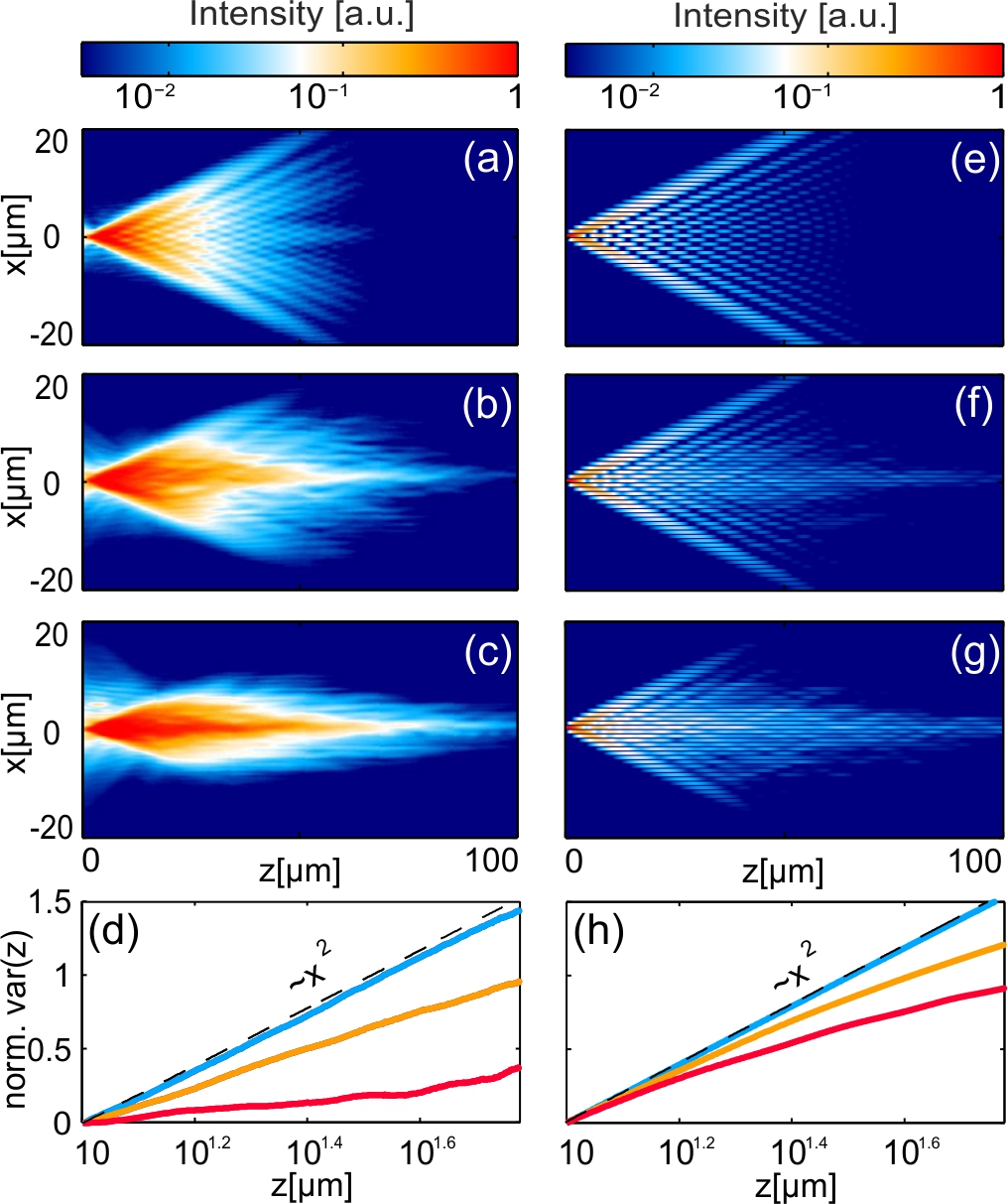}}
\caption{Real space intensity distribution averaged over up to 15 realizations: (a) $\Delta\beta/C=0.16$, (b) $\Delta\beta/C=1.2$, (c) $\Delta\beta/C=1.8$. Corresponding CMT simulations: (e-g). Averaged variance $\langle \rm{Var}(\textit{z})\rangle$ of the intensity patterns determined from the experimental results (d) as well as from the simulations (h). The blue line corresponds to $\Delta\beta/C=0.16$, the yellow line to $\Delta\beta/C=1.2$, the red line to $\Delta\beta/C=1.8$, and black dashed line indicates the discrete diffraction limit. }
\label{fig:RealSpace}
\end{figure}

The real-space leakage radiation micrograph of the array with nominally identical DLSPPWs  is shown in Fig.~\ref{fig:RealSpace} (a).  After excitation of the central DLSPPW, the SPP wave packet broadens into a cone with increasing propagation distance $z$ and a characteristic interference pattern occurs.
A large part of the intensity is carried in the two outer lobes. 
These features are signatures of ballistic transport \cite{eisenberg1998discrete}.
The effective coupling constant $C$ between adjacent DLSPPWs determines the opening angle of the cone.  
For the chosen geometrical parameters, we find $C=\SI{0.25}{\micro\metre}^{-1}$. In combination with the fabrication related variation of propagation constants, we obtain the disorder parameter $\Delta \beta / C= 0.16$ for this array. 

Figures~\ref{fig:RealSpace} (b) and (c) show  real-space leakage radiation micrographs of arrays with disorder parameter $\Delta\beta/C=1.2$ and $\Delta\beta/C=1.8$, respectively, averaged in each case over 15 different realizations.  
In the case of medium disorder ($\Delta\beta/C=1.2$), the wave packet is still spreading. However, a larger part of the intensity is concentrated in the center of the array. Finally, for strong disorder ($\Delta\beta/C=1.8$), the intensity is mostly confined in the central region of the array. That means that we have excited highly localized modes which is a definitive indication of transverse Anderson localization.

The averaged variance $\langle\rm{Var}(\textit{z})\rangle$ of the intensity pattern as a function of propagation distance $z$ provides a quantitative measure for the localization of the SPP wave packets~\cite{naether2013experimental}. For each disorder realization we calculate the variance
\begin{equation*}
{\mathrm{Var}}(z) = \frac{1}{I_{tot}(z)}\sum_{x} x^{2}\cdot I(x,z)-\left(\frac{1}{I_{tot}(z)}\sum_{x} x\cdot I(x,z)\right)^{2},
\label{eq:var}
\end{equation*}
where ${I_{\mathrm{tot}}(z)=\sum_{x} I(x,z)}$ is the total intensity as a function of $z$.
$\langle\rm{Var}(\textit{z})\rangle$ is then obtained by averaging $\rm{Var}(\textit{z})$ over the different experimental realizations of the same degree of disorder (see Fig~\ref{fig:RealSpace}(d)).
Without disorder the wave transport is expected to be ballistic and the corresponding variance should be a quadratic function of the propagation distance~\cite{stutzer2013superballistic}:
\begin{equation*}
\langle \rm{Var}(\textit{z})\rangle \sim \textit{z}^{2}.
\label{eq:DisDiff}
\end{equation*} 
The measured variance of the array with nominally identical DLSPPWs agrees well with this prediction. 
In the case of medium disorder ($\Delta\beta/C=1.2$), $\langle\rm{Var}(\textit{z})\rangle$ grows slower than $\sim z^{2}$. This can be attributed to the onset of  localization. 
For large disorder ($\Delta\beta/C=1.8$), the average variance is almost constant which is a clear signature of transverse Anderson localization. The slight increase of $\langle\rm{Var}(\textit{z})\rangle$ for large values of $z$ is most likely caused by scattered laser light.

We compare our experimental results to numerical calculations based on the coupled mode theory (CMT)~\cite{lin2007spatial}. Within this model we restrict ourselves to constant next neighbor coupling.
The coupling constant~$C$ as well as propagation constants~$\beta$ have been chosen to match the corresponding values of the fabricated arrays for the three different degrees of disorder. The calculated normalized intensity distributions averaged over 15 realizations (see Fig.~\ref{fig:RealSpace} (e,f,g)) as well as the calculated averaged variance (see Fig.~\ref{fig:RealSpace} (d)) show similar trends with increasing disorder as the corresponding experiments. 
However, the localization is more pronounced for the latter.
We attribute this to the variation of waveguide heights along the propagation direction, the next nearest neighbor coupling and the slightly altering coupling constants between the adjacent waveguides due to fabricational reasons. Nevertheless, the rather simple CMT model reflects the main properties of the spatial intensity evolution. 

\begin{figure}[ttt]
\centering
\fbox{\includegraphics[width=\linewidth]{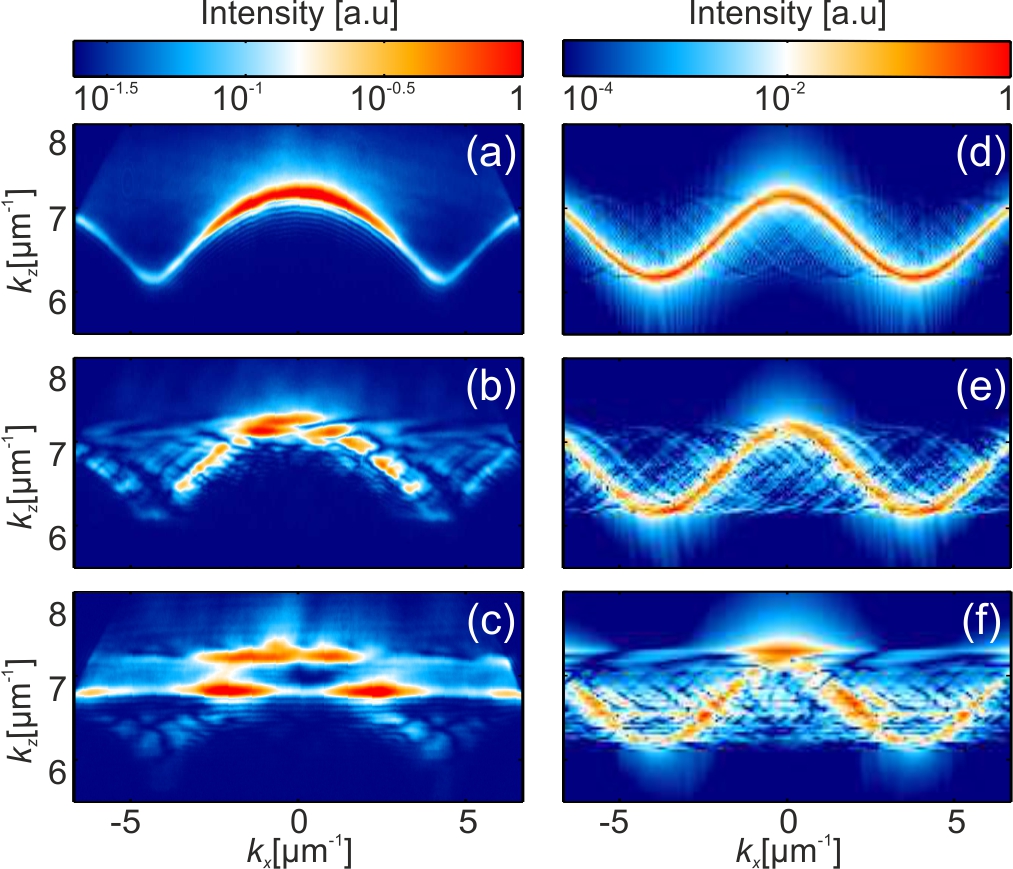}}
\caption{ Measured SPP momentum distributions for three different degrees of disorder (a) $\Delta\beta/C=0.16$, (b) $\Delta\beta/C=1.2$, (c) $\Delta\beta/C=1.8$.  Corresponding calculated SPP momentum distributions: (d) $\Delta\beta/C=0.16$, (e) $\Delta\beta/C=1.2$, (f) $\Delta\beta/C=1.8$.}
\label{fig:FourierSpace}
\end{figure}


It is of great interest to study the effect of disorder on the SPP momentum distribution $I(k_x,k_z)$. Based on the close relation between the temporal dynamics of electronic wave packets in solid state systems described by the tight-binding model and the corresponding spatial evolution of light in coupled waveguide arrays, 
we can interpret $k_x$ (transverse direction) as momentum and $k_{z}$ (SPP propagation direction) as energy.
Fig.~\ref{fig:FourierSpace}(a,b,c) show measured SPP momentum distributions for the three different degrees of disorder. Note that here the micrographs correspond to just one realization in each case. 
In the case of the array with nominally identical DLSPPWs ($\Delta\beta/C=0.16$), the momentum distribution consists of a single cosine-like band.
In view of the aforementioned relation between the tight-binding model and arrays of coupled waveguides, we can interpret this momentum distribution as the tight-binding band structure of a one-dimensional crystal with identical sites and identical couplings between the sites~\citep{christodoulides2003discretizing}.
Deviations of the measured band from an exact cosine-shape can most likely be attributed to next nearest neighbor coupling. 
With increasing disorder, we observe a transition from a continuous band ($\Delta\beta/C=0.16$) to a set of discrete modes that appear as straight lines with constant $k_z$ in the momentum distribution ($\Delta\beta/C=1.8$). This indicates the excitation of strongly localized modes and hence it is a further evidence of transverse Anderson localization. 
We note that for each disorder realization we observe a different set of localized modes. Although the qualitative features are however always the same. 

Figure \ref{fig:FourierSpace}(d,e,f) show calculated SPP momentum distributions for the three different degrees of disorder. They are obtained by two-dimensional Fourier transforms of the real space field distributions calculated with CMT.
The experimentally observed trends are qualitatively reproduced by the calculations. 
However, in accordance with the real-space data, we again find that the effects of disorder are less pronounced in the calculated SPP momentum distributions. 

In conclusion, we directly observed transverse Anderson localization of SPPs in disordered waveguide arrays using two different experimental approaches. In the first part of this work we demonstrated  the transition between ballistic wave packet expansion and transverse Anderson localization in real space. In the second part we presented the effect of disorder on the SPP momentum distribution. Our experimental results are in good agreement with CMT calculations.

We thank Mike Praemassing for help with the AFM measurements.

\bibliography{sample}


\end{document}